\documentclass[twocolumn,aps,bibtex,showpacs,amsmath,amssymb]{revtex4}
\usepackage{graphicx}% Include figure files
\usepackage{dcolumn}% Align table columns on decimal point
\usepackage{amsmath}%
\usepackage{amssymb}%

\usepackage{CJK}%

\begin{document}
\begin{CJK*}{GBK}{song}

%\preprint{APS/000-000}

\title{Derivation of Newton's law of gravitation based on a fluidic continuum model of vacuum and a sink flow model of particles}
% Force line breaks with \\

\author{Xiao-Song Wang}
%\altaffiliation[Also at ]{Physics Department, XYZ University.}
\email{wangxs2002@163.com}
%Lines break automatically or can be forced with \\
\affiliation{Institute of Mechanical and Power Engineering, Henan Polytechnic University,
Jiaozuo, Henan Province, 454000, China}
\date{\today}% It is always \today, today,
             %  but any date may be explicitly specified

\begin{abstract}
The main purpose of this paper is to seek a mechanical
interpretation of gravitational phenomena. We suppose that the
universe may be filled with a kind of fluid which may be called the
$\Omega (0)$ substratum. Thus, the inverse-square law of gravitation
is derived by methods of hydrodynamics based on a sink flow model of
particles. The first feature of this theory of gravitation is that
the gravitational interactions are transmitted by a kind of fluidic
medium. The second feature is the time dependence of gravitational
constant $G$ and gravitational mass. Newton's law of gravitation
is derived if we introduce an assumption that $G$ and the masses of
particles are changing so slowly that they can be treated as constants.
\end{abstract}

\keywords{gravitation; Newton's law of gravitation; inverse-square law; sink flow; aether; hydrodynamics.}

\maketitle

%\setcounter{section}{0} \stepcounter{section}
%%%%%%%%%%%%%%%%%%%%%%%%%%%%%%%%%%%%%%%%%%%%%%%%%%%%%%%%%%%%%%%%%%%%%%%%%%
\section{Introduction  \label{sec 100}}

\newtheorem{assumption}{\bfseries Assumption}

\newtheorem{definition}[assumption]{\bfseries Definition}

\newtheorem{lemma}[assumption]{\bfseries Lemma}

\newtheorem{proposition}[assumption]{\bfseries Proposition}

\newtheorem{theorem}[assumption]{\bfseries Theorem}

\newtheorem{wcorollary}[assumption]{\bfseries Corollary}

The Newton's law of gravitation can be written as
\begin{equation}\label{the Newton's law of gravitation}
\mbox{\upshape\bfseries{F}}_{21}=-\,G\,\frac{m_1m_2}{r^2}\,
\hat{\mbox{\upshape\bfseries{r}}}_{21}\,,
\end{equation}
where $m_1$ and $m_2$ are the masses of two particles, $r$ is the
distance between the two particles, $G$ is the gravitational
constant, $\mbox{\upshape\bfseries{F}}_{12}$
 is the force exerted on the particle with mass $m_2$
 by the particle with mass $m_1$,  $\hat{\mbox{\upshape\bfseries{r}}}_{21}$
denotes the unit vector directed outward along the line from the
particle with mass $m_1$ to the particle with mass $m_2$.

  The main purpose of this paper is to derive the Newton's law of
gravitation by means of fluid mechanics based on sink flow model of
particles.

  The motive of this paper is to seek a mechanism of gravitational
phenomena. The reasons why new models of gravity are interesting may
be summarized as follows.

   Firstly, there exists some astronomical phenomena that
could not be interpreted by the present theories of gravitation, for
instance, the Titius-Bode law \cite{Nietro1972}. New theories of
gravity may view these problems from new angles.

   Secondly, whether the gravitational constant $G$ depends on
time and space is still unknown
\cite{Gilvarry-Muller1972,Tomaschitz2000,Gaztanaga2002,Copi2004,Khoury-Weltman2004,Nagata2004,Biesiada-Malec2004}.
It is known that the gravitational constant $G$ is a constant in
 the Newton's theory of gravitation and in theory of general relativity.

   Thirdly, the mechanism of the action-at-a-distance gravitation remains an
unsolved problem in physics for more than 300 years
\cite{Whittaker1953,Hesse1961,Jaakkola1996}. Although theory of
general relativity is a field theory of gravity\cite{Fock1959}, the
concept of field is different from that of continuum mechanics
\cite{Truesdell1966,Fung1977,Eringen1980,Landau-Lifshitz1987}
because of the absence of a continuum in theory of general
relativity. Thus, theory of general relativity can only be regarded
as a phenomenological theory of gravity.

   Fourthly, we do not have a satisfactory quantum theory of gravity
presently
\cite{Carlip2001,Amelino-Camelia2002,Ahluwalia2002,Mitrofanov2003,Christian2005}.
One of the challenges in theoretical physics is to reconcile quantum
theory and theory of general relativity
\cite{Carlip2001,Collins2004}. New theories of gravity may open new
ways to solve this problem.

   Fifthly, one of the puzzles in physics is the problem
of dark matter and dark energy
\cite{Ellis2003,Linder2004,Bernardeau2003,Bergstrom2005,Beacom2005,Akerib2006,Feng2006,Fayet2006,Carena2006}.
New theories of gravity may provide new methods to attack this
problem\cite{Linder2004,Bernardeau2003}.

   Finally, we do not have a successful unified field theory presently.
Great progress has been made towards an unification of the four
fundamental interactions in the universe in the 20th century.
However, gravitation is still not unified successfully. New theories
of gravity may shed some light on this puzzle.

   To conclude, it seems that new considerations on gravitation is
needed. It is worthy keeping an open mind with respect to all the
theories of gravity before the above problems been solved.

   Now let us briefly review the long history of mechanical
interpretations of gravitational phenomena.
   Many philosophers and scientists, such as Laozi\cite{Laozi1995}, Thales,
Anaximenes, believed that everything in the universe is made of a
kind of fundamental substance\cite{Whittaker1953}. Descartes was the
first to bring the concept of aether into science by suggesting that
it has mechanical properties\cite{Whittaker1953}.
  Since the Newton's law of gravitation was published in
1687\cite{Newton-a}, this action-at-a-distance theory was criticized
by the French Cartesian\cite{Whittaker1953}. Newton admitted that
his law did not touch on the mechanism of
gravitation\cite{Cohen1980}. He tried to obtain a derivation of his
law based on Descartes' scientific research program \cite{Newton-a}.
Newton himself even suggested an explanation of gravity based on the
action of an aetherial medium pervading the
space\cite{Newton-b,Cohen1980}. Euler attempted to explain gravity
based on some hypotheses of a fluidic aether\cite{Whittaker1953}.

     In a remarkable paper published in 1905,
Einstein abandoned the concept of aether\cite{Einstein1905}.
However, Einstein's assertion did not cease the explorations of
aether
\cite{Whittaker1953,Vigier1980,Barut1988,Oldershaw1989a,Oldershaw1989b,
Carvalho2003,Arminjon2003,Carvalho-Oliveira2003,Jacobson2004,Davies2005,Levin-Wen2006}.
Einstein changed his view later and introduced his new concept of
ether\cite{Einstein1920,Kostro2000}. I regret to admit that it is
impossible for me to mention all the works related to this field in
history.
 Adolphe Martin and Roy Keys\cite{Martin2005a,Martin2005b,Martin-Keys1994} proposed a
gas model of vacuum to explain the physical
phenomena such as electromagnetism, gravitation, quantum mechanics
and the structure of elementary particles.

Inspired by the aforementioned thoughts and
others\cite{Lagally1922,Landweber-Yih1956,Yih1969,Faber1995,Currie2003},
we show that the Newton's law of gravitation is derived based on the
assumption that all the particles are made of singularities of a
kind of ideal fluid.

During the preparation of the manuscript, I noticed that John C.
Taylor had proposed an idea that the inverse-square law of
gravitation may be explained based on the concept of source or sink
\cite{Taylor2001}.

\section{Forces acting on sources and sinks in ideal fluids \label{sink}}
The purpose of this section is to calculate the forces between
sources and sinks in inviscid incompressible fluids which is called
ideal fluids usually.

Suppose the velocity field $\mbox{\upshape\bfseries{u}}$ of an ideal
fluid is irrotational, then we have
\cite{Lamb1932,Kochin1964,Yih1969,Wu1982a,Landau-Lifshitz1987,Faber1995,Currie2003},
\begin{equation}\label{velocity}
\mbox{\upshape\bfseries{u}}=\nabla\phi \,,
\end{equation}
where $\phi$ is the velocity potential, $\nabla =
\mbox{\upshape\bfseries{i}}\frac{\partial }{\partial x} +
\mbox{\upshape\bfseries{j}}\frac{\partial }{\partial y} +
\mbox{\upshape\bfseries{k}}\frac{\partial }{\partial z}$ is the
Hamilton operator.

It is known that the equation of mass conservation  of an ideal
fluid becomes Laplace's equation
\cite{Lamb1932,Kochin1964,Yih1969,Wu1982a,Faber1995,Currie2003},
\begin{equation}\label{Laplace's equation}
\nabla^2\phi=0\,,
\end{equation}
where $\phi$ is velocity potential, $\nabla^2 = \frac{\partial^2
}{\partial x^2} + \frac{\partial^2 }{\partial y^2} +
\frac{\partial^2 }{\partial z^2}$ is the Laplace operator.

Using spherical coordinates$(r,\theta,\varphi)$, a general form of
solution of Laplace's equation (\ref{Laplace's equation}) can be
obtained by separation of variables as\cite{Currie2003}
\begin{equation}\label{solution of Laplace's equation}
\phi (r,\theta)=\sum^{\infty}_{l=0}\left(
A_lr^l+\frac{B_l}{r^{l+1}}\right)P_l(\cos \theta)\,,
\end{equation}
where $A_l$ and $B_l$ are arbitrary constants,
 $P_l(x)$ are Legendre's function of the first kind which is defined as
\begin{equation}\label{Legendre's function}
P_l(x)=\frac{1}{2^ll!}\frac{\mathrm{d}^l}{\mathrm{d}x^l}\,(x^2-1)^l.
\end{equation}

If there exists a velocity field which is continuous and finite at
all points of the space, with the exception of individual isolated
points, then these isolated points are called singularities usually.

\begin{definition}\label{source or sink}
Suppose there exists a singularity  at point $P_0=(x_0,y_0,z_0)$.
 If the velocity field of the singularity at point $P=(x,y,z)$ is
\vspace*{-8pt}
\begin{equation}\label{velocity field of source or sink}
\mbox{\upshape\bfseries{u}}(x,y,z,t)=\frac{Q}{4\pi
r^2}\,\hat{\mbox{\upshape\bfseries{r}}}\,,
\end{equation}
where $r=\sqrt{(x-x_0)^2+(y-y_0)^2+(z-z_0)^2}$,
$\hat{\mbox{\upshape\bfseries{r}}}$
 denotes the unit vector directed outward along the line
from the singularity to the point $P=(x,y,z)$, then we call this
singularity a source if $Q>0$ or a sink if $Q<0$. $Q$ is called the
strength of the source or sink.
\end{definition}

Suppose a static point source with strength $Q$ locates at the
origin $(0, 0, 0)$. In order to calculate the volume leaving the
source per unit time, we may enclose the source with an arbitrary
spherical surface $S$ with radius $a$. A calculation shows that
\begin{equation}\label{volume leaving the source 2-10}
\int\hspace{-1.95ex}\int_{S} \hspace{-3.35ex}\bigcirc \ \
\mbox{\upshape\bfseries{u}}\cdot\mbox{\upshape\bfseries{n}} dS
 = \int\hspace{-1.95ex}\int_{S}
\hspace{-3.35ex}\bigcirc \ \ \frac{Q}{4\pi
a^2}\,\hat{\mbox{\upshape\bfseries{r}}}\cdot\mbox{\upshape\bfseries{n}} dS
= Q\,,
\end{equation}
where $\mbox{\upshape\bfseries{n}}$ denotes the unit vector directed
outward along the line from the origin of the coordinates to the
field point\linebreak $(x,y,z)$. Equation (\ref{volume leaving the source
2-10}) shows that the strength $Q$ of a source or sink evaluates the
volume of the fluid leaving or entering a control surface per unit
time.

From (\ref{solution of Laplace's equation}), we see that the
velocity potential $\phi (r,\theta)$ of a source or sink is a
solution of Laplace's equation (\ref{Laplace's equation}).

\begin{theorem}\label{force exerted on sources or sinks by fluids}
Suppose (1) there exists an ideal fluid  (2) the ideal fluid is
irrotational and barotropic, (3) the density $\rho$ is homogeneous,
that is
\begin{math}
\partial\rho / \partial x{=}\partial\rho / \partial y
{=}\partial\rho / \partial z{=}\partial\rho / \partial t\,{=}\,0\,,
\end{math}
(4) there are no external body forces exerted on the fluid, (5)the
fluid is unbounded and the velocity of the fluid at the infinity is
approaching to zero. Suppose a source or sink is stationary and is
immersed in the ideal fluid. Then, there is a force
\vspace*{-2pt}
\begin{equation}\label{force on the source in theorem}
\mbox{\upshape\bfseries{F}}_Q= -\,\rho
 Q\mbox{\upshape\bfseries{u}}_0
\end{equation}
exerted on the source by the fluid, where $\rho$ is the density of
the fluid, $Q$ is the strength of the source or the sink,
$\mbox{\upshape\bfseries{u}}_0$ is the velocity of the fluid at the
location of the source induced by all means other than the source
itself.
\end{theorem}

\mbox{\upshape\upshape\bfseries{Proof.}}
   Only the proof of the case of a source is needed.
Let us select the coordinates that is attached to the static fluid
at the infinity.

   We set the origin of the coordinates at the location of the source.
We surround the source by two arbitrary spherical surfaces $S_{\varepsilon}$ and $S$ centered at the origin of the coordinates. The radii of the surfaces $S_{\varepsilon}$ and $S$ are $r_{\varepsilon}$ and $r$ respectively. The radius $r_{\varepsilon}$ of the surfaces $S_{\varepsilon}$ is smaller than $r$ and is arbitrarily small. The outward unit normal to the surface $S$ is denoted by $\mbox{\upshape\bfseries{n}}$.

   Let $\tau(t)$ denotes the mass system of fluid enclosed in the volume
between the surface $S_{\varepsilon}$ and $S$ at time $t$.
   Let $\mbox{\upshape\bfseries{F}}_Q$ denotes the hydrodynamic force
exerted on the source by the mass system $\tau$, then a reaction $-\mbox{\upshape\bfseries{F}}_Q$ of
this force $\mbox{\upshape\bfseries{F}}_Q$ must act on the the surfaces $S_{\varepsilon}$ enclosing the source. Let $\mbox{\upshape\bfseries{F}}_S$ denotes the hydrodynamic
force exerted on the mass system $\tau$ due to the pressure
distribution on the surface $S$.

As an application of the Newton's second law of motion to the mass
system $\tau$, we have
\begin{equation}\label{the Newton's second law of motion}
\frac{\mathrm{D}\mbox{\upshape\bfseries{K}}}{\mathrm{D}t}=-\mbox{\upshape\bfseries{F}}_Q+\mbox{\upshape\bfseries{F}}_S\,,
\end{equation}
where $\mbox{\upshape\bfseries{K}}$
denotes momentum of the mass system $\tau$, $\mathrm{D} / \mathrm{D}t$ represents the material derivative
in the lagrangian system
\cite{Lamb1932,Kochin1964,Yih1969,Wu1982a,Landau-Lifshitz1987,Faber1995,Currie2003}.

The expressions of the momentum $\mbox{\upshape\bfseries{K}}$ and
the force $\mbox{\upshape\bfseries{F}}_S$ are
\begin{equation}\label{volume integral and surface integral}
\mbox{\upshape\bfseries{K}}
  =  \int\hspace{-1.5ex}\int\hspace{-1.5ex}\int_\tau \
\rho\,\mbox{\upshape\bfseries{u}} dV\,, \quad
\mbox{\upshape\bfseries{F}}_S
 =  \int\hspace{-1.95ex}\int_{S} \hspace{-3.35ex}\bigcirc \ \ (-p) \mbox{\upshape\bfseries{n}} dS\,,
\end{equation}
where the first integral is volume integral, the second integral is
surface integral,
 $\mbox{\upshape\bfseries{n}}$ denotes the unit vector directed outward
along the line from the origin of the coordinates to the field
point$(x,y,z)$.

Since the velocity field is irrotational, we have the following
relation
\begin{equation}
\mbox{\upshape\bfseries{u}}=\mbox{\upshape\bfseries{$\nabla$}} \phi\,,
\end{equation}
where $\phi$ is the velocity potential.

According to Ostrogradsky--Gauss theorem (see, for instance,
\cite{Kochin1964,Yih1969,Wu1982a,Faber1995,Currie2003}), we have
\begin{equation}\label{Ostrogradsky-Gauss}
\int\hspace{-1.5ex}\int\hspace{-1.5ex}\int_{\tau} \
\rho\,\mbox{\upshape\bfseries{u}} dV
=\int\hspace{-1.5ex}\int\hspace{-1.5ex}\int_{\tau} \
\rho\,\mbox{\upshape\bfseries{$\nabla$}} \phi  dV
=\int\hspace{-1.95ex}\int_{S} \hspace{-3.35ex}\bigcirc \ \
\rho\,\phi \mbox{\upshape\bfseries{n}} dS\,.
\end{equation}

Note that the mass system $\tau$ does not include the singularity
 at the origin.
 According to Reynolds' transport theorem
\cite{Kochin1964,Yih1969,Wu1982a,Faber1995,Currie2003}, we have
\begin{equation}\label{Reynolds' transport theorem}
\frac{\mathrm{D}}{\mathrm{D}t}\int\hspace{-1.5ex}\int\hspace{-1.5ex}\int_\tau \ \rho\mbox{\upshape\bfseries{u}} dV
=\frac{\partial}{\partial
t}\int\hspace{-1.5ex}\int\hspace{-1.5ex}\int_V
\rho\mbox{\upshape\bfseries{u}} dV +\int\hspace{-1.95ex}\int_{S}
\hspace{-3.35ex}\bigcirc \
\rho\mbox{\upshape\bfseries{u}} (\mbox{\upshape\bfseries{u}}\cdot\mbox{\upshape\bfseries{n}}) dS ,
\end{equation}
where $V$ is the volume fixed in space which coincide with the mass
system $\tau(t)$ at time $t$, that is $V=\tau(t)$.

Then, using (\ref{Reynolds' transport theorem}) , (\ref{volume
integral and surface integral}) and (\ref{Ostrogradsky-Gauss}), we
have
\begin{equation}\label{material derivative of K}
\frac{\mathrm{D}\mbox{\upshape\bfseries{K}}}{\mathrm{D}t}
=\int\hspace{-1.95ex}\int_{S} \hspace{-3.35ex}\bigcirc \ \
\rho\,\frac{\partial\phi}{\partial t}\,\mbox{\upshape\bfseries{n}} dS
 +\int\hspace{-1.95ex}\int_{S}\hspace{-3.35ex}\bigcirc \ \ \rho\mbox{\upshape\bfseries{u}} (\mbox{\upshape\bfseries{u}}\cdot\mbox{\upshape\bfseries{n}}) dS\,.
\end{equation}

According to Lagrange--Cauchy integral
\cite{Kochin1964,Yih1969,Wu1982a,Faber1995,Currie2003}, we have
\begin{equation}\label{Lagrange-Cauchy integral}
\frac{\partial\phi}{\partial t}+
\frac{(\mbox{\upshape\bfseries{$\nabla$}}\phi)^2}{2}+\frac{p}{\rho}=f(t)\,,
\end{equation}
where $f(t)$ is an arbitrary function of time $t$.

According to the assumption that the velocity $\mbox{\upshape\bfseries{u}}$ of the fluid at the infinity
is approaching to zero, we have
\begin{equation}\label{infinity 200-1600}
\mid \mbox{\upshape\bfseries{u}} \mid \rightarrow 0,  \quad r \rightarrow \infty.
\end{equation}

Noticing (\ref{solution of Laplace's equation}),
 $\phi (t)$ must be of the following form
\begin{equation}
\phi (r,\theta,t)=\sum^{\infty}_{l=0} \frac{B_l(t)}{r^{l+1}}\,P_l (\cos
\theta)\,,
\end{equation}
where $B_l(t), l \geqslant 0$ are functions of time $t$. Thus, we have
the following estimations at the infinity of the velocity field
\vspace*{-2pt}
\begin{equation}\label{estimation}
\phi =O\left( \frac{1}{r}\right), \quad \frac{\partial\phi}{\partial
t} =O\left( \frac{1}{r}\right), \quad r \rightarrow \infty\,,
\end{equation}
where $\varphi(x) = O(\psi(x)), x \rightarrow a$ stands for
$\overline{\lim}_{x \rightarrow a} \,{\mid}\varphi(x){\mid}\, /$\linebreak $\psi(x) = k,
(0 \leqslant k < +\infty).$

Applying (\ref{Lagrange-Cauchy integral}) at the infinity and using
(\ref{estimation}), we have
\begin{equation}\label{infinity 200-1800}
\frac{\partial\phi}{\partial t} \rightarrow 0, \quad p \rightarrow p_{\infty}, \quad r \rightarrow \infty,
\end{equation}
where $p_{\infty}$ is a constant.

Putting (\ref{infinity 200-1800}) and (\ref{infinity 200-1600}) into (\ref{Lagrange-Cauchy integral}), we have
\begin{equation}\label{infinity 200-1900}
f(t)=\frac{p_\infty}{\rho}.
\end{equation}

Putting (\ref{infinity 200-1900}) into (\ref{Lagrange-Cauchy integral}), we have
\begin{equation}\label{the pressure distribution on the surface S}
p=p_\infty-\rho\frac{\partial\phi}{\partial t}-
\frac{\rho(\mbox{\upshape\bfseries{u}}\cdot\mbox{\upshape\bfseries{u}})}{2}\,.
\end{equation}

Using (\ref{volume integral and surface integral}) and (\ref{the
pressure distribution on the surface S}), we have
\begin{equation}\label{force due to the pressure distribution on the surface S}
\mbox{\upshape\bfseries{F}}_S =\int\hspace{-1.95ex}\int_{S}
\hspace{-3.35ex}\bigcirc \ \ \rho\frac{\partial\phi}{\partial
t}\,\mbox{\upshape\bfseries{n}} dS
 +\int\hspace{-1.95ex}\int_{S} \hspace{-3.35ex}\bigcirc \ \
\frac{\rho(\mbox{\upshape\bfseries{u}}\cdot\mbox{\upshape\bfseries{u}})\,\mbox{\upshape\bfseries{n}}}{2}\,dS\,.
\end{equation}

Using (\ref{the Newton's second law of motion}), (\ref{material
derivative of K}), (\ref{force due to the pressure distribution on
the surface S}), we have
\begin{equation}\label{force on the source a}
\mbox{\upshape\bfseries{F}}_Q =\int\hspace{-1.95ex}\int_{S}
\hspace{-3.35ex}\bigcirc \ \
\left[\frac{1}{2}\,\rho(\mbox{\upshape\bfseries{u}}\cdot\mbox{\upshape\bfseries{u}})\mbox{\upshape\bfseries{n}}
-\rho\mbox{\upshape\bfseries{u}}(\mbox{\upshape\bfseries{u}}\cdot\mbox{\upshape\bfseries{n}})\right]dS\,.
\end{equation}

Now let us calculate this velocity $\mbox{\upshape\bfseries{u}}$ in
order to obtain $\mbox{\upshape\bfseries{F}}_Q$. Since the velocity
field induced by the source $Q$ is (\ref{velocity field of source or
sink}), then according to the superposition principle
 of velocity field of ideal fluids, the velocity on the surface $S$ is
\begin{equation}\label{velocity on the surface S}
\mbox{\upshape\bfseries{u}} =\frac{Q}{4\pi
r^2}\mbox{\upshape\bfseries{n}}+\mbox{\upshape\bfseries{u}}_0,
\end{equation}
where $\mbox{\upshape\bfseries{n}}$ denotes the unit vector directed
outward along the line from the origin of the coordinates to the
field point\linebreak $(x,y,z)$. Using (\ref{force on the source a}) and
(\ref{velocity on the surface S}), we have
\begin{eqnarray}\label{force on the source b}
\mbox{\upshape\bfseries{F}}_Q
 &=& \rho  \int\hspace{-1.95ex}\int_{S} \hspace{-3.35ex}\bigcirc
 \;  \left [ \frac{Q^2}{32\pi^2 r^4}\,\mbox{\upshape\bfseries{n}}
 +\frac{1}{2}(\mbox{\upshape\bfseries{u}}_0\cdot\mbox{\upshape\bfseries{u}}_0)\,\mbox{\upshape\bfseries{n}}\,\right.\nonumber \\
 &&\left.-\frac{Q}{4\pi r^2}\,\mbox{\upshape\bfseries{u}}_0
 -(\mbox{\upshape\bfseries{u}}_0\cdot\mbox{\upshape\bfseries{n}})\,\mbox{\upshape\bfseries{u}}_0
 \right]dS\,.
\end{eqnarray}

Since the radius $r$ can be arbitrarily small, the velocity
$\mbox{\upshape\bfseries{u}}_0$ can be treated as a constant in the
integral of (\ref{force on the source b}). Thus, (\ref{force on the
source b}) turns out to be
\begin{equation}\label{force on the source c}
\mbox{\upshape\bfseries{F}}_Q = -\rho\int\hspace{-1.95ex}\int_{S}
\hspace{-3.35ex}\bigcirc \ \ \frac{Q}{4\pi
r^2}\,\mbox{\upshape\bfseries{u}}_0 \,dS\,.
\end{equation}

Since again $\mbox{\upshape\bfseries{u}}_0$ can be treated as a
constant, (\ref{force on the source c}) turns out to be (\ref{force
on the source in theorem}). This completes the proof.  $\Box$

\mbox{\upshape\upshape\bfseries{Remark.}} Lagally \cite{Lagally1922},
 Landweber and Yih \cite{Landweber-Yih1956,Yih1969}, Faber \cite{Faber1995} and
Currie \cite{Currie2003} obtained the same result of Theorem
\ref{force exerted on sources or sinks by fluids} for the special
case where the velocity field is steady.

  Theorem \ref{force exerted on sources or sinks by fluids} only considers the situation
that the sources or sinks are at rest. Now let us consider the case
that the sources or sinks are moving in the fluid.

\begin{theorem}\label{force exerted on
moving sources or sinks by fluids} Suppose the presuppositions
(1), (2), (3), (4) and (5) in Theorem \ref{force exerted on sources or
sinks by fluids} are valid and a source or a sink is moving in the
fluid with a velocity $\mbox{\upshape\bfseries{v}}_s$, then there is
a force
\begin{equation}\label{force on moving source in theorem}
\mbox{\upshape\bfseries{F}}_Q= -\rho\,
 Q\,(\mbox{\upshape\bfseries{u}}_f-\mbox{\upshape\bfseries{v}}_s)
\end{equation}
is exerted on the source by the fluid, where $\rho$ is the density
of the fluid, $Q$ is the strength of the source or the sink,
$\mbox{\upshape\bfseries{u}}_f$ is the velocity of the fluid at the
location of the source induced by all means other than the source
itself.
\end{theorem}

\mbox{\upshape\upshape\bfseries{Proof.}}
 The velocity of the fluid relative to the source at the
location of the source is
$\mbox{\upshape\bfseries{u}}_f-\mbox{\upshape\bfseries{v}}_s$. Let
us select the coordinates that is attached to the source and set the
origin of the coordinates at the location of the source. Then
(\ref{force on moving source in theorem}) can be arrived following
the same procedures in the proof of Theorem \ref{force exerted on
sources or sinks by fluids}.     $\Box$

Applying Theorem \ref{force exerted on moving sources or sinks by
fluids} to the situation
 that a source or sink is exposed to the velocity field of another source or sink,\linebreak we have:

\begin{wcorollary}\label{singularity-singularity forces}
Suppose the presuppositions (1), (2), (3), (4) and (5) in Theorem
\ref{force exerted on sources or sinks by fluids} are valid and a
source or a sink with strength $Q_2$ is exposed to the velocity
field of another source or sink with strength $Q_1$, then the force
$\mbox{\upshape\bfseries{F}}_{21}$
 exerted on the singularity with strength $Q_2$
 by the velocity field of the singularity with strength $Q_1$ is
\vspace*{-3pt}
\begin{equation}\label{force exerted on singularity Q_2 by singularity Q_1}
\mbox{\upshape\bfseries{F}}_{21}=-\rho Q_2\,\frac{Q_1}{4\pi
r^2}\,\hat{\mbox{\upshape\bfseries{r}}}_{21} + \rho
Q_2\mbox{\upshape\bfseries{v}}_2\,,
\end{equation}
where $\hat{\mbox{\upshape\bfseries{r}}}_{21}$ denotes the unit
vector directed outward along the line from the singularity with
strength $Q_1$ to the singularity with strength $Q_2$, $r$ is the
distance between the two singularities,
$\mbox{\upshape\bfseries{v}}_2$ is the velocity of the source with
strength $Q_2$.
\end{wcorollary}

\section{Derivation of inverse-square-law of gravitation \label{gravitation}}
Since quantum theory shows that vacuum is not empty and has physical
effects, e.g., the Casimir
effect\cite{Lamoreaux2005,Intravaia-Lambrecht2005,Guo-Zhao2004,Davies2005},
it is valuable to probe vacuum by introducing the following
hypotheses:

\begin{assumption}\label{Omega (1)}
Suppose the universe is filled by an ideal fluid named $\Omega (0)$
substratum; the ideal fluid fulfil the conditions (2), (3), (4), (5)
in Theorem \ref{force exerted on sources or sinks by fluids}.
\end{assumption}

This fluid may be named $\Omega (0)$ substratum in order to
distinguish with Cartesian aether.

The idea that all microscopic particles are source or sink flows in a fluidic substratum is not new. For instance, in order to compare fluid motions with electric fields, J. C. Maxwell introduced an analogy between source or sink flows and electric charges (\cite{Whittaker1951}, p243). H. Poincar$\acute{e}$ also speculates that matters are holes in fluidic aether (\cite{PoincareH1997}, p171). A. Einstein and L. Infeld said (\cite{EinsteinAInfeldL}, p256-257):"{\itshape Matter is where the concentration of energy is great, field where the concentration of energy is small. $\cdots$ What impresses our senses as matter is really a great concentration of energy into a comparatively small space. We could regard matter as the regions in space where the field is extremely strong.}" It seems that they are suggesting that particles are some kinds of singularities of field.

Following these researchers, we adopt the idea that particles may be looked as
singularities in fields. Noticing
(\ref{force exerted on singularity Q_2 by singularity Q_1}), it is
nature to introduce the following:
\begin{assumption}\label{monad}
All the microscopic particles were made up of a kind of elementary
sinks of $\Omega (0)$ substratum. These elementary sinks were
created simultaneously. The initial masses and the strengths of the
elementary sinks are the same.
\end{assumption}

We may call these elementary sinks as monads.
Suppose a particle with mass $m$ is composed of $N$ monads. Then,
according to Assumption \ref{monad}, we have:
\begin{eqnarray}
&&\rule{-1.5cm}{0pt} m_0(t) =m_0(0)+\rho q_0 t\,, \label{the mass of a monad is increasing 1} \\
&&\rule{-1.5cm}{0pt} Q  = -\,N q_0\,, \quad m(t) = N m_0(t) = - \,\frac{Q}{q_0}\,m_0(t)\,,
\label{the mass of a monad is increasing 2} \\[-3pt]
 &&\rule{-1.5cm}{0pt} \frac{dm_0}{dt} =\rho q_0\,, \quad \frac{dm}{dt}  = -\rho Q\,,\label{the mass of a monad is increasing 3}
\end{eqnarray}
where $m_0(t)$ is the mass of monad at time $t$, $-q_0( q_0 > 0)$ is
the strength of a monad, $m(t)$ is the mass of a particle at time
$t$,
 $Q$ is the strength of the particle,
 $N$ is the number of monads that make up the particle,
 $\rho$ is the density of the $\Omega (0)$ substratum, $t\geqslant 0$.

From (\ref{the mass of a monad is increasing 3}), we see that the
mass $m_0$ of a monad is increasing since $q_0$
 evaluates the volume of the $\Omega (0)$ substratum fluid entering the monad per unit time.
From (\ref{the mass of a monad is increasing 3}), we also see that
the mass of a monad or a particle is increasing linearly.

 Based on
Assumption \ref{Omega (1)} and Assumption \ref{monad}, the motion of
a particle is determined by:
\begin{theorem}\label{motion of a particle}
The equation of motion of a particle is
\vspace*{-2pt}
\begin{equation}\label{the equation of motion of a particle}
m(t)\,\frac{\mathrm{d}\mbox{\upshape\bfseries{v}}}{\mathrm{d}t}
=\frac{\rho q_{0}}{m_{0}(t)}\, m(t) \mbox{\upshape\bfseries{u}}
-\frac{ \rho q_{0}}{m_{0}(t)}\, m(t) \mbox{\upshape\bfseries{v}}
+\mbox{\upshape\bfseries{F}}\,,
\end{equation}
where $m_0(t)$ is the mass of monad at time $t$, $-q_0$ is the\linebreak
strength of a monad, $m(t)$ is the mass of a particle at time $t$,
$\mbox{\upshape\bfseries{v}}$ is the velocity of the particle,
$\mbox{\upshape\bfseries{u}}$ is the velocity of the $\Omega (0)$
substratum at the location of the particle induced by all means
other than the particle itself, $\mbox{\upshape\bfseries{F}}$
denotes other forces.
\end{theorem}

\mbox{\upshape\upshape\bfseries{Proof.}}
%\begin{proof}
 Applying the Newton's second law and
Theorem \ref{force exerted on moving sources or sinks by fluids} to
this particle, we have
\begin{equation}\label{Newton 300-3000}
m\frac{d \mbox{\upshape\bfseries{v}}}{d t} = -\rho
Q(\mbox{\upshape\bfseries{u}}-\mbox{\upshape\bfseries{v}})
+\mbox{\upshape\bfseries{F}}.
\end{equation}

Putting (\ref{the mass of a monad is increasing 2}) into (\ref{Newton 300-3000}), we get (\ref{the equation of motion of a particle}).
$\Box$

Formula (\ref{the equation of motion of a particle}) shows that there exists
a universal damping force
\vspace*{-2pt}
\begin{equation}\label{universal damping force}
\mbox{\upshape\bfseries{F}}_d = -\frac{\rho
q_0}{m_0}\, m \mbox{\upshape\bfseries{v}}
\end{equation}
exerted on each particle.

Now let us consider a system consists of two particles. Based on
Assumption \ref{monad}, applying Theorem \ref{motion of a particle}
to this system, we have:

\begin{wcorollary}\label{motion of a system consists of two particles}
Suppose there is a system consists of two particles and there are no
other forces exerted on the particles, then the equations of motion
of this system are
\vspace*{-2pt}
\begin{eqnarray}\label{equations of motion of a system consists of two particles}
&&\hspace{-1.5cm} m_1 \frac{d\mbox{\upshape\bfseries{v}}_1}{dt}
  =  -\frac{\rho q_0}{m_0}\, m_1 \mbox{\upshape\bfseries{v}}_1
 -\frac{\rho q^2_0}{4\pi m^2_0} \frac{m_1m_2}{r^2}\,
\hat{\mbox{\upshape\bfseries{r}}}_{12} \label{equations of motion of a system consists of two particles 1}\\
&&\hspace{-1.5cm} m_2 \frac{d\mbox{\upshape\bfseries{v}}_2}{dt}
  =  -\frac{\rho q_0}{m_0}\,m_2\mbox{\upshape\bfseries{v}}_2
 -\frac{\rho q^2_0}{4\pi
 m^2_0}\frac{m_1m_2}{r^2}\,\hat{\mbox{\upshape\bfseries{r}}}_{21}\,,\label{equations of motion of a system consists of two particles 2}
\end{eqnarray}
where $m_{i=1, 2}$ is the mass of the particles,
$\mbox{\upshape\bfseries{v}}_{i=1, 2}$ is the velocity of the
particles, $m_0$ is the mass of a monad,
 $-q_0$ is the strength of a monad,
 $\rho$ is the density of the $\Omega (0)$ substratum,
$\hat{\mbox{\upshape\bfseries{r}}}_{12}$ denotes the unit vector
directed outward along the line from the particle with mass $m_2(t)$
to the particle with mass $m_1(t)$,
$\hat{\mbox{\upshape\bfseries{r}}}_{21}$ denotes the unit vector
directed outward along the line from the particle with mass $m_1(t)$
to the particle with mass $m_2(t)$.
\end{wcorollary}

 Ignoring the damping forces in (\ref{equations of motion of a system consists of two particles 2}), we have:
\begin{wcorollary}\label{Inverse-Square-Law of Gravitation}
Suppose
(1)~$\mbox{\upshape\bfseries{v}}_{i=1, 2}\ll\mbox{\upshape\bfseries{u}}_{i=1, 2}$,
where $\mbox{\upshape\bfseries{v}}_i$ is the velocity of the
particle with mass $m_i$, $\mbox{\upshape\bfseries{u}}_i$ is the
velocity of the $\Omega (0)$ substratum at the location of the
particle with mass $m_i$ induced by the other particle,
 (2)~there are no other forces exerted on the particles,
then the force $\mbox{\upshape\bfseries{F}}_{21}(t)$
 exerted on the particle with mass $m_2(t)$
 by the velocity field of $\Omega (0)$ substratum induced by the particle with mass $m_1(t)$ is
 \begin{equation}\label{the inverse-square-law of gravitation}
\mbox{\upshape\bfseries{F}}_{21}(t)=- G(t)\,\frac{m_1(t) m_2(t)}{r^2}\,
\hat{\mbox{\upshape\bfseries{r}}}_{21}\,,
\end{equation}
where
\begin{equation}\label{gravitational constant 5-10}
G(t)=\frac{\rho  q^2_0}{4\pi m^2_0(t)}\,,
\end{equation}
$\hat{\mbox{\upshape\bfseries{r}}}_{21}$ denotes the unit vector
directed outward along the line from the particle with mass $m_1(t)$
to the particle with mass $m_2(t)$, $r$ is the distance between the
two particles.
\end{wcorollary}

Corollary \ref{Inverse-Square-Law of Gravitation} is similar to
Newton's inverse-square-law of gravitation (\ref{the Newton's
law of gravitation}) except for two differences. The first
difference is that $m_{i=1,2}$ are constants in the Newton's law
(\ref{the Newton's law of gravitation}) while in Corollary
\ref{Inverse-Square-Law of Gravitation} they are functions of time
$t$. The second difference is that $G$ is a constant in the Newton's
theory while here $G$ is a function of time $t$.

Let us now introduce an assumption that $G$ and the masses of
particles are changing so slowly relative to the time scale of
human beings that they can be treated as constants approximately.
Thus, the Newton's law of gravitation may be considered as a result
of Corollary \ref{Inverse-Square-Law of Gravitation} based on this
assumption.

\section{Superposition principle of gravitational field \label{superposition}}
The definition of gravitational field $\mbox{\upshape\bfseries{g}}$
of a particle with mass $m$ is
\begin{equation}\label{field 400-100}
\mbox{\upshape\bfseries{g}}=\frac{\mbox{\upshape\bfseries{F}}}{m_{test}},
\end{equation}
where $m_{test}$ is the mass of a test point mass,
$\mbox{\upshape\bfseries{F}}$ is the gravitational force exerted on
the test point mass by the gravitational field of the particle with
mass $m$.

Based on (\ref{field 400-100}), Theorem \ref{motion of a particle} and Corollary
\ref{Inverse-Square-Law of Gravitation}, we have
\begin{equation}\label{gravitational field}
\mbox{\upshape\bfseries{g}}=\frac{\rho
q_0}{m_0}\,\mbox{\upshape\bfseries{u}}\,,
\end{equation}
where $\rho$ is the density of the $\Omega (0)$ substratum, $m_0$ is
the mass of a monad, $q_0$ is the strength of a monad,
$\mbox{\upshape\bfseries{u}}$ is the velocity of the $\Omega (0)$
substratum at the location of the test point mass induced by the
particle mass $m$.

From (\ref{gravitational field}), we see
that the superposition principle of gravitational field is deduced
from the superposition theorem of the velocity field of ideal
fluids.

\section{Equations of gravitational field of continuously distributed
particles \label{field}}
The definition of the volume density of continuously
distributed sink is
\begin{equation}\label{definition of the volume density of continuously distributed sink}
\rho_s=\lim_{\triangle V \rightarrow 0}\frac{\triangle Q}{\triangle
V},
\end{equation}
where $\triangle V$ is a small volume, $\triangle Q$ is the sum of
strengthes of all the sinks in the volume $\triangle V$.

Now let us to derive the continuity equation of the taothe
$\Omega (0)$ substratum from the
principle of mass conservation. Consider an arbitrary volume $V$
bounded by a closed surface $S$ and fixed in space. Suppose there
are some point sources or sinks continuously distributed in the
volume $V$. The total mass in volume $V$ is
\begin{equation}
M=\int\hspace{-1.5ex}\int\hspace{-1.5ex}\int_{V} \rho dV,
\end{equation}
where $\rho$ is the density of the $\Omega (0)$ substratum.

The rate of increase of the total mass in volume $V$ is
\begin{equation}\label{The rate of increase of the total mass}
 \frac{\partial M}{\partial t}=\frac{\partial }{\partial t}\int\hspace{-1.5ex}\int\hspace{-1.5ex}\int_{V}\rho
 dV.
\end{equation}

The rate of mass outflow through the surface $S$ is
\begin{equation}\label{The rate of mass outflow 5-10}
 \int\hspace{-1.9ex}\int_{S}\hspace{-3.27ex}\bigcirc\rho (\mbox{\bfseries{u}} \cdot \mbox{\bfseries{n}} )dS,
\end{equation}
where $\mbox{\bfseries{u}}$ is the velocity field of the $\Omega (0)$ substratum.

The rate of mass created inside the volume $V$ is
\begin{equation}\label{the rate of mass created}
\int\hspace{-1.5ex}\int\hspace{-1.5ex}\int_{V} \rho\rho_{s}dV.
\end{equation}

Now according to the principle of mass conservation, and making use
of (\ref{The rate of increase of the total mass}), (\ref{The rate of
mass outflow 5-10})
 and  (\ref{the rate of mass created}), we have
\begin{equation}\label{mass conservation 5-10}
 \frac{\partial }{\partial t}\int\hspace{-1.5ex}\int\hspace{-1.5ex}\int_{V} \rho dV
 = \int\hspace{-1.5ex}\int\hspace{-1.5ex}\int_{V} \rho\rho_{s} dV - \int\hspace{-1.9ex}\int_{S}\hspace{-3.27ex}\bigcirc\rho (\mbox{\bfseries{u}} \cdot \mbox{\bfseries{n}} )dS
\end{equation}

According to Ostrogradsky--Gauss theorem (refer to, for instance,
\cite{Kochin1964,Yih1969,Wu1982a,Faber1995,Currie2003}), we have
\begin{equation}\label{The rate of mass outflow 5-20}
 \int\hspace{-1.9ex}\int_{S}\hspace{-3.27ex}\bigcirc\rho (\mbox{\bfseries{u}} \cdot \mbox{\bfseries{n}}
 )dS = \int\hspace{-1.5ex}\int\hspace{-1.5ex}\int_{V} \nabla \cdot (\rho \mbox{\bfseries{u}})dV,
\end{equation}

 Using (\ref{The rate of mass outflow 5-20}), (\ref{mass conservation 5-10}) becomes
\begin{equation}\label{mass conservation 5-20}
 \frac{\partial }{\partial t}\int\hspace{-1.5ex}\int\hspace{-1.5ex}\int_{V} \rho dV
 = \int\hspace{-1.5ex}\int\hspace{-1.5ex}\int_{V} \rho\rho_{s} dV - \int\hspace{-1.5ex}\int\hspace{-1.5ex}\int_{V} \nabla \cdot (\rho \mbox{\bfseries{u}})dV
\end{equation}

 Since the volume $V$ is arbitrary, from (\ref{mass conservation 5-20}) we have
\begin{equation}\label{equation of mass conservation 5-30}
 \frac{\partial \rho}{\partial t} + \nabla \cdot (\rho \mbox{\bfseries{u}}) = \rho\rho_{s}.
\end{equation}

Since the $\Omega (0)$ substratum is homogeneous, i.e.,
\begin{math}
\partial\rho/\partial x=\partial\rho/\partial y
 =\partial\rho/\partial z=\partial\rho/\partial t=0,
\end{math}
(\ref{equation of mass conservation 5-30}) becomes
\begin{equation}\label{equation of mass conservation 5-40}
 \nabla \cdot  \mbox{\bfseries{u}} = \rho_{s}.
\end{equation}

Thus, from (\ref{gravitational field}) and (\ref{equation of mass
conservation 5-40}), we have
\begin{equation}
 \nabla \cdot\mbox{\upshape\bfseries{G}} = \frac{\rho
q_0\rho_s}{m_0}.
\end{equation}

Let $\rho_m$ denotes the volume mass density of continuously
distributed particles. According to Assumption \ref{monad}, and
using (\ref{the mass of a monad is increasing 2})and
(\ref{definition of the volume density of continuously distributed
sink}), we have
\begin{equation}
\rho_m = - \frac{m_0 \rho_s}{q_0}.
\end{equation}
Thus, noticing that the velocity field $\mbox{\upshape\bfseries{u}}$
of the $\Omega (0)$ substratum is irrotational and $G=\rho q^2_0/(4\pi m^2_0)$, the
equations of gravitational field of continuously distributed
particles can be summarized as
\begin{equation}\label{equations of gravitational field}
\left\{
\begin{array}{ll}
\nabla \times\mbox{\upshape\bfseries{G}}
 =\mbox{\upshape\upshape\bfseries{0}},\\
\nabla \cdot\mbox{\upshape\bfseries{G}}
 =-4\pi G \rho_m.
\end{array}
\right.
\end{equation}

\section{The equivalence of inertial mass and gravitational mass \label{mass}}
According to Assumption \ref{monad} and Corollary \ref{Inverse-Square-Law of Gravitation}, we have
\begin{equation}\label{mass 100}
m_{inertial}=m_{gravitational},
\end{equation}
where $m_{inertial}$ is the inertial mass of a particle,
$m_{gravitational}$ is the gravitational mass of the particle.

\section{Time dependence of gravitational constant $G$ and mass}\label{dependence 800}
The time dependence of the gravitational constant $G$ can be seen from (\ref{gravitational constant 5-10}).

The time dependence of the gravitational mass can be seen from (\ref{the mass of a monad is increasing 3}).

\section{Possible space dependence of gravitational constant $G$}\label{dependence 900}
If the density $\rho$ of the $\Omega (0)$
substratum varies from place to place, i.e., $\rho=\rho(\mbox{\upshape\bfseries{r}})$, then the space dependence of  the gravitational constant $G$ can be seen from (\ref{gravitational constant 5-10}).

\section{Discussion \label{sec 1000}}
Although the new formula of gravitation (\ref{the inverse-square-law of gravitation}) is similar to Newton's inverse-square-law of gravitation (\ref{the Newton's law of gravitation}), there exists the following five differences between this theory and Newton's theory.

1. The gravitational masses are constants in Newton's law, while in (\ref{the inverse-square-law of gravitation}) they are functions of time $t$.

2. The gravitational constant $G$ is a constant in Newton's theory, while in (\ref{gravitational constant 5-10}) $G$ depends on time $t$.

3. In this theory, the parameter $G$ depends on the density $\rho_{\Omega(0)}$ of the $\Omega (0)$
substratum. If $\rho_{\Omega(0)}$ varies from place to place, then the space dependence of the gravitational constant $G$ can be seen from (\ref{gravitational constant 5-10}).

4. In Newton's theory, the gravity is action-at-a-distance \cite{Whittaker1953}. In this theory, the gravity is transmitted by the $\Omega (0)$ substratum.

5. Newton's law of gravitation is an assumption. In this theory, (\ref{the inverse-square-law of gravitation}) is derived by methods of classical fluid mechanics based on some assumptions.

\section{Conclusion \label{sec 1100}}
We suppose that the universe may be filled with a kind of fluid
which may be called the $\Omega (0)$ substratum. Thus, the
inverse-square law of gravitation is derived by methods of
hydrodynamics based on a sink flow model of particles. There are two
features of this theory of gravitation. The first feature is that
the gravitational interactions are transmitted by a kind of fluidic
medium. The second feature is the time dependence of gravitational
constant and gravitational mass. Newton's law of gravitation is
derived if we introduce an assumption that $G$ and the masses of
particles are changing so slowly that they can be treated as
constants. As a byproduct, it is shown that there exists a universal
damping force exerted on each particle.

\section*{Acknowledgments \label{sec 1200}}
I am grateful to Robert L. Oldershaw for informing me his researches in the field of
 Self-Similar Cosmological Paradigm (SSCP) and discrete fractal cosmological models during the preparation of the manuscript. I wish to express my thanks to Dr. Roy Keys for providing me three
 interesting articles \cite{Martin2005a,Martin2005b,Martin-Keys1994}.

%\section*{Appendix \label{sec 1300}}

%\end{CJK}{GBK}{song}
\end{CJK*}

\end{document}